\documentclass[12pt,preprint]{aastex}
\begin{document}

\newcommand{\kms}{$\,\mbox{km}\,\mbox{s}^{-1}$}
\newcommand{\kmsfour}{$\,\mbox{km}^{-4}\,\mbox{s}^{4}\,
	{\cal M}_{\odot}$}
\newcommand{\kmsML}{$\,\mbox{km}\,\mbox{s}^{-1}\, {\cal M}_{\odot}^{-1/2}\,
	L_{\odot}^{1/2}$}
\newcommand{\kmskpc}{\ensuremath{\;{\mathrm{km}^2\,
	\mathrm{s}^{-2}\,\mathrm{kpc}^{-1}}}}
\newcommand{\etal}{et al.}
\newcommand{\LCDM}{$\Lambda$CDM}
\newcommand{\ML}{\ensuremath{\Upsilon_{\star}}}
\newcommand{\amax}{\ensuremath{a_{\dagger}}}
\newcommand{\Om}{\ensuremath{\Omega_m}}
\newcommand{\OL}{\ensuremath{\Omega_{\Lambda}}}
\newcommand{\Q}{\ensuremath{{\cal Q}}}
\newcommand{\Lsun}{\ensuremath{L_{\odot}}}
\newcommand{\Msun}{\ensuremath{{\cal M}_{\odot}}}
\newcommand{\mass}{\ensuremath{{\cal M}}}
\newcommand{\Vst}{\ensuremath{V_{\star}}}
\newcommand{\vst}{\ensuremath{v_{\star}}}
\newcommand{\Vg}{\ensuremath{V_{g}}}
\newcommand{\Vb}{\ensuremath{V_{b}}}
\newcommand{\Vh}{\ensuremath{V_h}}
\newcommand{\gn}{\ensuremath{g_N}}
\newcommand{\So}{\ensuremath{\Sigma_0}}
\newcommand{\Sg}{\ensuremath{\Sigma_g}}
\newcommand{\Sb}{\ensuremath{\Sigma_{b}}}
\newcommand{\Sst}{\ensuremath{\Sigma_{\star}}}
\newcommand{\HI}{H{\sc i}\ }


\title{The Mass Discrepancy-Acceleration Relation: \\
         Disk Mass and the Dark Matter Distribution}

\author{Stacy S.~McGaugh} 

\affil{Department of Astronomy, University of Maryland}
\affil{College Park, MD 20742-2421}    
\email{ssm@astro.umd.edu}

\begin{abstract}
The mass discrepancy in disk galaxies is shown to be well correlated
with acceleration, increasing systematically with decreasing acceleration
below a critical scale $a_0 \approx 3700\kmskpc
= 1.2 \times 10^{-10}\;\mathrm{m}\,\mathrm{s}^{-2}$.  For each galaxy,
there is an optimal choice of stellar mass-to-light ratio which minimizes
the scatter in this mass discrepancy-acceleration relation.  
The same mass-to-light ratios also minimize the scatter in the
baryonic Tully-Fisher relation and are in excellent
agreement with the expectations of stellar population synthesis.
Once the disk mass is determined in this fashion, 
the dark matter distribution is specified.
The circular velocity attributable to the dark matter
can be expressed as a simple equation which depends only on
the observed distribution of baryonic mass.
It is a challenge to understand how this very fine-tuned coupling 
between mass and light comes about.  
\end{abstract}


\section{Introduction}

The masses of stellar disks and the distribution of mass in dark matter
halos pose a coupled problem.
Rotation curves provide good measures of the mass enclosed by disks.
But it has been difficult to disentangle how much of this mass is in
the stellar disk, and how much is in the dark matter halo.  
Consequently, both the mass of the stellar disk and
the distribution of the dark matter, $\rho(r)$, have been unclear. 

There have long been suggestions of a close connection between mass and light
in spiral galaxies.  Perhaps the most obvious is the Tully-Fisher
relation (Tully \& Fisher 1977).  
Beyond this global scaling relation, there are indications of a local
coupling between mass and light (e.g., Rubin \etal\ 1985; Bahcall \&
Casertano 1985; Persic \& Salucci 1991).  One manifestation of this is in the
efficacy of maximum disk (e.g., van Albada \& Sancisi 1986) in describing
the inner parts of rotation curves.  If one scales up the stellar contribution
to the rotation curves of high surface brightness (HSB) spirals to the maximum
allowed by the data, one often finds a good match in the details (the
``bumps and wiggles'') between the shape of the rotation curve and that 
predicted by the observed stellar mass (e.g., Kalnajs 1983; Sellwood 1999; 
Palunas \& Williams 2000).  This only works out to some radius where dark
matter must be invoked, but does suggest that the preponderance of the mass
at small radii is stellar.  Beyond that, it is often possible to scale up
the gas component to explain the remainder of the rotation curve
(Hoekstra, van Albada, \& Sancisi 2001).  Moreover, while some dark
matter may be needed to stabilize disks, detailed analyses
of disk stability frequently require rather heavy disks in order to
drive the observed bars and spiral features (e.g., Athanassoula, Bosma, \& 
Papaioannou 1987; Debattista \& Sellwood 1998, 2000; Weiner, Sellwood, \& 
Williams 2001; Fuchs 2003a; Bissantz, Englmaier, \& Gerhard 2003;
Kranz, Slyz, \& Rix 2003).

While these lines of evidence favor nearly maximal disks in HSB
spirals, there are contradictory indications as well.  The most significant
of these is the lack of surface brightness (or scale length) residuals in
the Tully-Fisher relation (Sprayberry \etal\ 1995; Zwaan \etal\ 1995;
Hoffman \etal\ 1996; Tully \& Verheijen 1997).
The apparent lack of influence of the distribution of disk mass on the
Tully-Fisher relation suggests that disks are submaximal
(McGaugh \& de Blok 1998a; Courteau \& Rix 1999).
However, galaxies which occupy the same location in the Tully-Fisher
plane can have very different rotation curve shapes (de Blok \& McGaugh
1996; Tully \& Verheijen 1997).  This excludes the simple hypothesis that
galaxies of equal luminosity reside in identical halos with no significant
influence from the disk.

Recent data for low surface brightness (LSB) disks complicate matters further.
These objects show large mass discrepancies down to small radii, implying
that they are dark matter dominated with very submaximal disks
(de Blok \& McGaugh 1997).  However, it is often formally possible to
obtain a fit with something like a traditional maximum disk (where the
peak velocity of the disk component is comparable to $V_{flat}$),
albeit at the cost of absurdly high ($> 10\; \Msun/\Lsun$) mass-to-light
ratios (de Blok \& McGaugh 1997; Swaters, Madore, \& Trewhella 2000; 
McGaugh, Rubin, \& de Blok 2001).
As anticipated by McGaugh \& de Blok (1998b), density wave analyses imply
nearly maximal disks for LSB galaxies (Fuchs 2002, 2003b). 
These high mass-to-light ratios are unlikely for stellar populations,
so one might consider a disk component of dark matter in addition to the
usual halo.  This seems contrived, and also causes problems with the
baryonic Tully-Fisher relation (McGaugh \etal\ 2000; Bell \& de Jong 2001).  
This relation between mass and rotation velocity works best, in the
sense of having minimal scatter, for disk masses which are consistent with
stellar population mass-to-light ratios (\S 3).
Maximal disks in LSB galaxies increase the scatter in the baryonic
Tully-Fisher relation.  

Among these apparently contradictory lines of evidence, there is
nevertheless a clear theme.  The luminous and dark components
are intimately linked.  This is true not only in a global sense, but
also in a local one.
This might be paraphrased as Renzo's Rule: {\it ``For any feature in the 
luminosity profile there is a corresponding feature in the rotation curve''}
(Sancisi 2003).  The distribution of baryonic 
mass is completely predictive of the distribution
of dark matter, even in dark matter dominated LSB galaxies.

Renzo's rule is an empirical statement which is mathematically 
encapsulated by MOND (Milgrom 1983).  MOND is a modified force
law hypothesized as an alternative to dark matter, and remains a 
viable possibility (Sanders \& McGaugh 2002).  Even if dark matter
is correct (as is widely presumed), MOND is still useful
as a compact description of the mass discrepancy in spirals
(Sanders \& Begeman 1994).

In this paper, I show that the stellar mass-to-light ratios determined from
MOND fits to rotation curves are optimal in a purely Newtonian sense.
I derive a simple expression for the corresponding dark matter distribution, 
and generalize this to apply for {\it any\/} choice of stellar mass-to-light ratio.  
This expression provides a simple yet stringent test for dark matter
theories which seek to explain rotation curves.

\section{The Data}

The data used here are from the 
sample of disk galaxies collected by Sanders \& McGaugh (2002)
where a complete Table of galaxy properties can be found.
This represents the accumulated work of a good many (and many good) people,
including Begeman (1987), Begeman, Broeils, \& Sanders (1991), Broeils (1992),
Sanders (1996), de Blok, McGaugh, \& van der Hulst (1996), de Blok (1997),
Verheijen (1997), Sanders \& Verheijen (1998),
McGaugh \& de Blok (1998a,b), de Blok \& McGaugh (1998),
Verheijen (2001), and Verheijen \& Sancisi (2001).
This is a wonderful compilation of information for investigating the
details of mass distributions in spiral galaxies.

The data for each galaxy includes the rotation curve and the distribution
of the observed mass components.  These include the stellar disk
and bulge (if present), and the \HI gas component.   Rotation curves
have been derived from velocity fields of nearby galaxies.
The distribution of stellar mass has been derived from a variety
of passbands (see Sanders \& McGaugh 2002 and references therein).  
Much of the sample has $K'$-band photometry (Tully \etal\ 1996)
which provides the closest mapping between light and stellar mass. 
Other bands
(usually $R$ or $B$) go deeper, and provide consistent results, albeit
with a larger scatter in the inferred mass-to-light ratio.  The \HI data are
the most time consuming to obtain, and are the limiting factor on the size
of the  sample. The \HI distribution is nevertheless an essential ingredient, 
as the atomic gas often dominates the baryonic mass surface density at 
large radii.

Two examples of the data are shown in Fig.~1.
The rotation attributable to stars and gas is shown for both galaxies, 
as is the total rotation due to baryons:
\begin{equation}
\Vb^2 = \Vst^2 + \Vg^2.
\end{equation} 
These components of the velocity are derived from the observed distribution
of surface brightness $I(r)$ through numerical inversion of the Poisson 
equation:  
\begin{equation}
\nabla^2 \Phi_b = 4 \pi G \rho_b.
\end{equation}
For disks,
$\rho_b = \Sigma(r) \delta(z)$ where $\Sigma(r)$ is the azimuthally averaged
radial surface mass distribution and $\delta(z)$ is the mass distribution
perpendicular to the disk.  The vertical mass distribution
makes little difference to the mass models unless the
disk is very thick (axis ratios $<$ 5:1; e.g., de Blok \etal\ 2003); thin disks
are assumed here.  A more important factor is the mass-to-light ratio \ML\ 
which relates the stellar mass distribution to the observed light distribution 
[$\Sigma(r) = \ML I(r)$].  The mass-to-light ratio is assumed to be 
constant with radius\footnote{From a population
perspective, one would expect a modest gradient in \ML\ for typical color
gradients.  There are hints of this in the dynamical data, but this is a 
subtle effect compared to the mean value of \ML.}
so that the observed distribution of light $I(r)$ specifies the quantity \vst:
$\Vst^2(r) = \ML \vst^2(r)$.  (Note that \vst\ has units \kmsML.)
The velocity due to the atomic gas component is more
directly related to the observed 21 cm flux, with the correction 
$\mass_g = 4/3 \mass_{\mathrm{HI}}$ for helium and metals.
Molecular and ionized gas are assumed to make a negligible contribution
to the baryonic mass budget.  This is almost certainly fair for ionized gas,
and in most cases for molecular gas (e.g., Olling 1996). 
Molecular gas closely follows the distribution of star light (Regan \etal\ 2001), 
so it is subsumed\footnote{In effect, $\mass_{\star}$ represents all components
that follow the stellar distribution: stars plus molecular gas plus whatever else
might lurk in the disk.  So if, for example, I give $\ML = 1.1\;\Msun/\Lsun$ 
but it is later determined that the molecular gas mass is 10\% of 
the stellar mass, then this would mean that the stars alone would have 
$\ML = 1.0\;\Msun/\Lsun$.} into the stellar mass-to-light ratio. 

The centripetal acceleration predicted by 
Newtonian gravity for the observed baryonic mass components is
\begin{equation}
g_N = \frac{\Vb^2}{r} = \left| \frac{\partial \Phi_b}{\partial r} \right|.
\end{equation}
This is related to the observed centripetal acceleration produced
by all mass components
\begin{equation}
a = \frac{V^2}{r} = \left| \frac{\partial \Phi}{\partial r} \right|
\end{equation}
by MOND through
\begin{equation}
\mu(x) a = g_N,
\end{equation}
where $x = a/a_0$ and $a_0$ is a constant.
The value of $a_0$ is taken to be the same in all cases,
$a_0 = 1.2 \times 10\;{\mathrm m}\,{\mathrm s}^{-2}$
(Begeman \etal\ 1991). 
The interpolation function $\mu(x)$ commonly used
in rotation curve fitting is
\begin{equation}
\mu(x) = \frac{x}{\sqrt{1+x^2}}
\end{equation}
(Milgrom 1983; Sanders \& McGaugh 2002).  This is effectively
just a scaling of the velocity due to the baryonic component.  There
is a simple formula for mapping $\Vb(r)$ to the total velocity $V(r)$ (Fig.~1).  
This procedure has only a single fitting parameter, the stellar mass-to-light ratio
\ML.  The physical origin of this scaling aside, the simple fact that it works
provides a very useful constraint on the problem.

The MOND procedure has been successfully applied to $\sim 100$
galaxies.  Two dozen of these cases have yet
to be published (Swaters \& Sanders 2004, in preparation), leaving a
useful sample of 74 galaxies (Sanders \& McGaugh 2002).
This is a very large sample for data of this quality and extent.

Though not a 
complete sample in terms of a survey, this is not important to the purpose
here.  What is important is coverage of the parameter space over which
disk galaxies exist.  The sample galaxies 
cover a large range in rotation velocity ($V_{flat} = 50$ to 300\kms),
luminosity ($L_B = 5 \times 10^7$ to $2 \times 10^{11} \Lsun$),
scale length ($h = 0.5$ to 13 kpc), central surface brightness
($\mu_0^B = 19.6$ to 24.2 mag.\ arcsec$^{-2}$), and gas mass fraction
($f_g = \mass_g/(\mass_{\star}+\mass_g) = 0.07$ to 0.95).  This range covers 
most of the parameter space over which disk galaxies are known to reside.  

The efficacy of MOND in fitting rotation curves
is well established (Sanders \& McGaugh 2002).  The quality of these
fits is illustrated by Figs.~2 and 3, which show the global rms and local
velocity residuals of the MOND fits.
The rms residuals are small, $< 10\%$ in most cases (Fig.~2a).
Of course, some data are more accurate than others, so it is interesting
to ask how MOND performs as the data improve.  To this end, we restrict
the sample to include only those points within each galaxy for which the
formal uncertainty on the velocity is better than 5\%.  In crude terms, this
is roughly the level at which various astrophysical effects limit one's
knowledge of the true circular velocity (see discussions in McGaugh \etal\ 
2001 and de Blok \etal\ 2001).  This restriction removes the low accuracy
points from the rotation curves of individual galaxies, and in 14 galaxies
no data remain.  The rms residuals improve after the less accurate data
are rejected (Fig.~2b): the best data are fit best by MOND.  

In the analysis, it is assumed that the orbits being traced are circular.
This assumption must fail at some level, causing a deviation of the
observed rotation curve from the circular velocity curve of the potential.
This effect will usually be most severe at small radii where the
gradient in the rotation curve is large, and where the velocity dispersion may
contribute significantly to the total kinematic budget, especially in systems
with large bulges.  The assumption of circular motion will cause the
MOND-predicted velocities to exceed the observed ones by a modest amount.
This is precisely what is seen in Fig.~3(a), where there is a ``beard'' of  points
at small radii with $\Delta V = V-V_{\mathrm{MOND}} < 0$.  
This effect is consistent with
the modest amount of non-circular motion one would naturally expect for 
this sample.\footnote{In most of the studies from which the data are drawn, 
one selection criterion is a reasonable degree of axis-symmetry:  one should
not attempt to apply the assumption of circular motion to objects with
grossly asymmetric rotation curves.}

Two galaxies listed by Sanders \& McGaugh (2002) are not included here:
NGC 3198 and NGC 2841.  These galaxies have MOND fits which 
are sensitive to the distance measurement, as discussed in detail by
Bottema \etal\ (2002).  It is not surprising that this
should occasionally be an issue since it is acceleration which matters in 
MOND fits.  Since $a = V^2/r$, the acceleration can be uncertain (through
$r = \theta D$) even if $V$ is well measured.  NGC 3198 and NGC 2841 
either fall right on target in Fig.~2, and adhere to the same Tully-Fisher and
mass discrepancy-acceleration relations as the other 74 galaxies, or they 
are extreme outliers from these relations, depending on their true distances
(and other uncertainties, like the angle of the warp in NGC 2841).
If these objects can not be reconciled with MOND,
then they constitute a falsification of the theory.  Such a situation would not,
however, alter any of the conclusions drawn here in the context of dark matter.
From this perspective, they would merely be rare cases which deviate from
the norm.

\section{The Empirical Coupling between Mass and Light}

\subsection{The Acceleration Scale}

From a purely Newtonian perspective, the mass discrepancy in disk
galaxies sets in at a particular acceleration scale.  Whether this is also
true in other systems, especially rich clusters of galaxies, is less clear
(Aguirre, Schaye, \& Quataert 2001; Sanders 2003), so the discussion
here is restricted to rotationally supported disk galaxies.  In disks, it is
very clear that acceleration is the relevant physical scale (Fig.~4).

The ordinate of Fig.~4 plots the amplitude of the mass discrepancy
${\cal D}$, as quantified by the ratio of the gradient of the total
gravitational potential to that of the baryons:
\begin{equation}
{\cal D} \equiv \frac{\Phi'}{\Phi_b'} = \frac{V^2}{V_b^2}
 = \frac{V^2}{\ML \vst^2 + V_g^2}.
\end{equation}
In the limit of spherical mass distributions, this is equivalent to the
ratio of total to baryonic mass (McGaugh 1999).  The mass discrepancy
is plotted against several physical scales in Fig.~4:
radius, orbital frequency, and centripetal acceleration.  The data
for all 74 galaxies are plotted together, a total of 1,145 individual
resolved velocity measurements.  

With so many data from so many
different galaxies of such widely varying properties, one might naively
expect the result to be a scatter plot.  Indeed, this is essentially the
case when the mass discrepancy is plotted against radius (top panels 
of Fig.~4).  There are galaxies where the need for dark matter is not
apparent until quite large radii, and others in which the mass discrepancy
sets in already at small radii.  There is no particular radius at which the
mass discrepancy always occurs, as we would expect if the rotation 
curve of the dynamically dominant dark matter $\Vh(r)$ depends little 
on the minority baryons.

Much the same is true when the data are plotted against orbital frequency.
There is a hint of incipient organization, but there is also still a lot of scatter.
Yet when the mass discrepancy is plotted against acceleration, something 
remarkable happens.  All the data from all the galaxies align.  
The correlation is remarkably strong considering the number of galaxies
and independent measurements which went into it.

The baryonic potential computed for Fig.~4, while purely Newtonian, assumes
MOND mass-to-light ratios.  McGaugh (1999) showed that acceleration is
the physical scale with which the mass discrepancy correlates best irrespective 
of the choice of stellar mass-to-light ratio.  However, it is interesting to see
how the correlation is affected by other choices of \ML.

Fig.~5 shows the mass discrepancy-acceleration relation for four choices
of stellar mass-to-light ratio (\ML\ in equation~7).  Illustrated are 
maximum disk,  MOND, and two choices of stellar population models.
The choice labeled ``popsynth'' computes the mass-to-light ratio from the
model of Bell \etal\ (2003b):
\begin{eqnarray}
\log \ML^B = 1.737 (B-V)-0.942  \\
\log \ML^K = 0.135 (B-V)-0.206
\end{eqnarray}
(from their Table 7).  ``Half popsynth'' has the same color dependence, but to 
emulate a lightweight IMF has its normalization reduced 
by a factor of two.  This is chosen to represent
the lower end of the realm of plausibility (Kroupa 2002); reality is
presumably bracketed between this and maximum disk.  
Mass-to-light ratios estimated in this fashion will not be perfect of course,
but this does provide a useful estimate for what we expect for
stellar populations.

Colors for the
galaxies are taken from the original sources where available.  In other
cases, they are taken from NED,\footnote{This research has made use of the 
NASA/IPAC Extragalactic Database (NED) which is operated by the Jet 
Propulsion Laboratory, California Institute of Technology, under contract 
with the National Aeronautics and Space Administration.} with precedence
given to the RC3 (de Vaucouleurs \etal\ 1995) where available to maximize
uniformity.  $B-V$ colors were found for 50 of the galaxies; popsynth \ML\ 
are not computed for the remainder for lack of input to equations 8 and 9.

The mass discrepancy-acceleration relation is clear in Fig.~5 for all choices
of mass-to-light ratio.  It works nearly as
well for maximum disk as it does for MOND mass-to-light ratios.  
This is because high surface brightness galaxies are close to or in
the Newtonian regime ($a \gtrsim a_0$) at small radii, so the MOND
and maximum disk mass-to-light ratios are similar.\footnote{By construction,
MOND \ML\ must be maximal (i.e., no dark matter) in the limit $a \gg a_0$.
As a practical matter, accelerations never much exceed $a_0$ in galactic disks,
so the MOND \ML\ are always at least a bit less than maximum.} 
The popsynth
choice also works well.  Individual rotation curves can be perceived
in a few cases, but the number of deviant galaxies is small.  Only as the
mass-to-light ratio becomes implausibly small (half popsynth)
does the relation begin to fall apart (as it must in the absurd limit
$\ML \rightarrow 0$).

The scatter in the mass discrepancy-acceleration relation is minimized for
MOND mass-to-light ratios.  This is no accident.  The trend in the data in 
this figure is, in effect, the inverse of the MOND interpolation function 
$\mu(x)$ (equation 6).
The MOND mass-to-light ratios have been chosen to minimize the scatter
away from this function.  Nevertheless, Fig.~5 is a purely Newtonian
diagram.  We can explicitly fit the data in these diagrams to describe the
dependence of the mass discrepancy on acceleration with a function 
${\cal D}(x) = \mu^{-1}(x)$ (Fig.~4) 
or ${\cal D}(y) = \mu^{-1}[\mathrm{g}(y)] \approx \mu^{-1/2}(y)$ (Fig.~5)
where $y = g_N/a_0$.  Regardless of how we choose to represent it
mathematically, or whether we call it MOND or dark matter, this organization
is clearly present in the data.

It seems quite unlikely that this high degree of organization can be an accident.
Certainly it can not be a product of bad data.  Systematic errors could
obscure a real relation, but they can not conspire to give the appearance of
one where none exists.  Rather, the mass discrepancy-acceleration relation 
must be a real aspect of nature. 

\subsection{The Baryonic Tully-Fisher Relation}

In addition to the local mass discrepancy-acceleration relation, there also
exists the global Tully-Fisher relation.  Usually expressed as a relation between 
luminosity and linewidth, it has long been suspected that this is a manifestation
of some more fundamental relation between mass and rotation velocity
(e.g., Freeman 1999).  McGaugh \etal\ (2000) explicitly showed
that there is indeed a more fundamental relation between the 
baryonic mass (stars plus gas) and the rotation velocity 
(see also Bell \& de Jong 2001; McGaugh 2003).   For galaxies with
resolved rotation curves, the velocity in the flat part of the rotation curve,
$V_{flat}$, is the obvious quantity of interest, and provides an excellent
correlation (Fig. 6).

Fig.~6 illustrates the baryonic Tully-Fisher relation for the same choices of
mass-to-light ratio and data accuracy as in Fig.~5.  As in Fig.~5, there is a
clear correlation in all cases, though it begins to degrade for the 
half-popsynth case.  Also as in Fig.~5, the scatter is minimized for MOND
mass-to-light ratios.  

Verheijen (1997, 2001) notes that the Tully-Fisher relation for the Ursa Major
cluster is remarkably tight, with only barely room for the intrinsic scatter 
expected from stellar population variations, let alone that expected
from halo-to-halo variations.  Fig.~6 further emphasizes that 
whatever physical mechanism underpins the Tully-Fisher relation, it is a 
remarkably strong one with much less scatter than one would nominally 
expect (e.g., Eisenstein \& Loeb 1996; McGaugh \& de Blok 1998a).  
This tight relation is no surprise if required by the force law.  MOND requires 
a baryonic Tully-Fisher relation of the form
\begin{equation}
\mass_b = {\cal A} V_{flat}^4.
\end{equation}
A fit to the data in Fig.~6 does indeed find a slope of 4 with a  
normalization ${\cal A} = 50$\kmsfour.  This is somewhat higher than the
normalization of McGaugh \etal\ (2000), though consistent within the 
uncertainties.
The biggest difference is that here I have only used galaxies with $V_{flat}$
explicitly measured from resolved rotation curves as opposed to line
widths.

In MOND, the normalization of the baryonic Tully-Fisher relation is
${\cal A} = \chi/(a_0 G)$, where $\chi$ is a factor of order unity which
accounts for the fact that thin disks rotate faster than the equivalent 
spherical mass distribution.  Since the galaxies here have all been fit with
$a_0 = 1.2 \times 10\;\mathrm{m}\,\mathrm{s}^{-2}$,
we measure $<$$\chi$$>$ $= 0.79$, as expected (McGaugh \& de Blok 1998b).
In MOND, the baryonic Tully-Fisher relation should be perfect with no scatter.
That there remains a small amount of scatter in Fig.~6 is simply a reflection
of the residual uncertainties in the data.

Of the choices of mass-to-light ratio considered here,
there is no better choice than MOND for minimizing the
scatter in the baryonic Tully-Fisher relation.  
Maximum disk does a good job, but does less well in this respect because
low surface brightness galaxies tend to have very large maximum disk
mass-to-light ratios (de Blok \& McGaugh 1997; Swaters \etal\ 2000).
This pushes up the scatter as a wide range of mass is inferred at a given
luminosity.  It also pushes up the normalization at low mass, where most
galaxies are of low surface brightness (top panel of Fig.~6).  It is only
with very high mass-to-light ratios ($\ML \gtrsim 10\;\Msun/\Lsun$) that
the stellar mass is significant in these low mass galaxies; in the lower
three panels the gas mass dominates the baryonic total and these points
do not budge.

Presuming mass is indeed more fundamental than luminosity, one would
expect a good choice of population synthesis model mass-to-light ratios to
reduce the scatter below that observed in luminosity.  That is, a proper matching of
\ML\ to each galaxy would remove the component of the scatter due to it.
While this does happen with MOND, the popsynth choice
does not reduce the scatter (Bell \& de Jong 2001), rather increasing it
slightly.  In retrospect, this is probably due to the fact that one expects a
fair amount of scatter in the \ML-color relation (equations~8 and 9).  
So while the color may
be a good indicator of \ML\ in the mean, when applied to any particular
galaxy the scatter simply propagates.

\subsection{The Optimal Mass-to-Light Ratio}

Obtaining the mass of a spiral disk is simply a matter of choosing the
right mass-to-light ratio.  The MOND mass-to-light ratio appears to be 
optimal, even in a purely Newtonian sense, in that it
minimizes the scatter in both the global baryonic Tully-Fisher relation and
the local mass discrepancy-acceleration relation.  This can be further tested
against the expectations of stellar population synthesis models (Fig. 7).

The MOND mass-to-light ratios are in excellent agreement with population
synthesis models.  These dynamically determined \ML\ are completely
independent of the models to which they are compared.  Yet they agree
well in normalization with the best-guess IMF of Bell \& de Jong (2001).
In addition, the expected trend of \ML\ with color is apparent.
The $K'$-band mass-to-light ratio is almost independent of color, while
there is a clear trend in the $B$-band for redder galaxies to have higher
mass-to-light ratios.  The slope of this relation is consistent with the models,
and one can even see indications of the turndown in $\ML^B$ for 
$B-V < 0.5$ apparent in Fig.~B1 of Portinari, Sommer-Larsen, \& Tantalo 
(2004).  Moreover, the scatter
is larger in $B$ than in $K'$, again as expected.  In this respect, the
MOND mass-to-light ratios are more consistent with population synthesis
models than popsynth itself, as the latter applies equation~8 or 9 
without allowance for the expected scatter.

If we accept 
the agreement with population models at face value, we can begin
to place constraints on the IMF.
Bell \& de Jong (2001) argue for a ``scaled'' Salpeter IMF where the 
best mass-to-light ratio is scaled by a factor ${\cal X}$ from the familiar 
Salpeter IMF:  $\ML = {\cal X} \ML^{\mathrm{Salpeter}}$.  
By tracing the lower envelope of the
maximum disk data of Verheijen (1997) (lower left panel of Fig.~7),
Bell \& de Jong (2001) argue for ${\cal X} = 0.7$.  This is a good argument,
but one could also argue that one expects a fair amount of scatter 
about the mean population line, so that a good number of disks should
fall below this line even if all disks are maximal.  
If we take this attitude then ${\cal X} \approx 1.3$ provides
a good description of the data.  This is not particularly unreasonable 
from a population perspective, but there are a number of individual galaxies
where $\ML^{\mathrm{max}}$ is uncomfortably high 
(typically $\ML^B \approx 10\, \Msun/\Lsun$ for blue LSB galaxies).

For the optimal mass-to-light ratios from MOND, 
${\cal X} \approx 3/4$.
This is in quite good agreement with the detailed investigation of 
Kroupa \& Weidner (2003), who find $0.72 < {\cal X} < 0.83$
from direct integration of the observed IMF,
including brown dwarfs.  There is a slight tension between the mean of
the $B$ and $K'$ bands, in that the $B$-band prefers slightly higher ${\cal X}$.
This may be due to residual shortcomings in the models, or may be due
to the uncertainty in \ML\ caused by that in the distance to Ursa Major
(from whence come all the $K'$-band data).
The distance determined by Sakai \etal\ (2000) is rather larger than
that of Pierce \& Tully (1988), and this cluster is by far the largest
outlier from the mean\footnote{The Hubble constant determination
of Sakai \etal\ (2000) is 
$H_0 = 71\;\mathrm{km}\,\mathrm{s}^{-1}\,\mathrm{Mpc}^{-1}$ 
but their distance for the Ursa Major cluster implies
$H_0 = 53\;\mathrm{km}\,\mathrm{s}^{-1}\,\mathrm{Mpc}^{-1}$.}
in the Sakai \etal\ sample.  Hence the data still
require refinement before a definitive conclusion about the value of ${\cal X}$
can be made.  Nevertheless, it is clearly in the right neighborhood,
and extreme values, both high and low, can be discounted.  
For example, ${\cal X} > 1.5$ would
exceed maximum disk for the majority of $K'$-band cases.  
${\cal X} < 0.35$ (half-popsynth or less) would destroy the correlations 
which are clear in the data.

There are now three items which point to MOND mass-to-light ratios
being optimal in a purely Newtonian sense:  (1) they minimize the scatter
in the local mass discrepancy-acceleration relation, 
(2) they minimize the scatter in the global baryonic Tully-Fisher relation, 
and (3) they could hardly be in better agreement
with stellar population models.  The obvious conclusion is that these are 
in fact the correct mass-to-light ratios.  This would appear to solve the long
standing problem of the uncertainty in disk masses.  With the
disk mass specified, the dark matter distribution follows.

\section{The Dark Matter Distribution}

\subsection{Constraints on Halo Parameters from the Acceleration Scale}

The presence of an acceleration scale in the data provides strong constraints
on the parameters of halo models.  This follows simply from the observation
that there is a point (in Fig.~5) at which the mass discrepancy appears.
The constraints discussed here (in \S 4.1) do not depend on
the details of MOND fits or the coupling of mass and light below the
critical acceleration scale, but merely upon the fact that there is a
critical acceleration scale.

This relevance of the acceleration scale to dark matter halos has
been noted by Brada \& Milgrom (1999).  They pointed out
that one can only attribute to dark matter a maximum acceleration
not already accounted for by the stars.  They demonstrate that there
is a formal upper limit to the halo acceleration
\begin{equation}
a^{\mathrm{max}}_h \le \eta \amax,
\end{equation}
where \amax\ is the critical acceleration and $\eta$ is a numerical parameter.
For MOND mass-to-light ratios, $\amax\ \equiv a_0$.
The value of $\eta$ depends weakly on the adopted form of the MOND
interpolation function, $\mu(x)$.  This can be evaluated numerically 
(see Brada \& Milgrom 1999);
for the commonly used form of $\mu(x)$ (equation~6)
$\eta = 0.30$ which I adopt here.  The range of $\eta$
spanned by plausible interpolation functions is $0.25 \le \eta \le 0.37$, so
the particular choice of interpolation function is not very important.  

Phrasing the constraint in this way makes the relationship to MOND clear.
However, the constraint follows simply from the observation that there is an 
upper limit to the force being provided by the dark halo.  We can generalize
the MOND result to allow for any choice of mass-to-light ratio.  
For each choice of \ML, the effective value of \amax\ can be determined 
by fitting a function ${\cal D}(y)$ 
to each panel in Fig.~5.  The result is
\begin{itemize}
\item for maximum disk: $\amax \approx 2300\kmskpc$;
\item for popsynth: $\amax \approx 4000\kmskpc$; and
\item for half popsynth, $a_{\dagger} \approx 10,000\kmskpc$.
\end{itemize}

The critical acceleration for popsynth is indistinguishable from MOND: 
$\amax \approx a_0$.  For maximum disk,
$\amax < a_0$, which simply says that the mass discrepancy appears
at a somewhat lower acceleration, as should be true by construction.
A similar result can be seen in other data (e.g., Palunas \& Williams 2000),
though the availability of the gas component is necessary for the
smooth appearance in Fig.~5.  The correlation appears to break down
at low accelerations if the gas component is neglected, since it is often
the dominant baryonic component at large radii and low accelerations.
For half popsynth, $\amax > a_0$.
The fit is still tolerable, but the uncertainty in \amax\ begins to grow
large.  In the limit $\ML \rightarrow 0$, we would of course infer mass
discrepancies everywhere, so $\amax \rightarrow \infty$.

The scale \amax\ represent an upper limit on the acceleration which
can be caused by the dark matter halo.  This can be translated to limits
on halo parameters for any choice of halo model.  These will be rather
conservative limits, as we can only require that the maximum acceleration
produced by a halo model not exceed \amax.  However, there is no
guarantee that the maximum acceleration in any given galaxy will
actually occur at the radius of maximum acceleration of a particular
halo model.  In addition, there is no reason to expect that real galaxies
need ever exhibit accelerations as high as \amax. 
LSB galaxies have large mass discrepancies 
precisely because $V^2/r < a_0$ at all radii.  Probably the
dominant uncertainty is in the choice of \ML\ and hence the appropriate
value of \amax, though from \S 3 the most obvious choice is
$\amax = a_0$.  

In the following sections, I derive the limits on halo parameters
which follow from the acceleration scale limit for the most 
common halo models.  These are the constant density core 
pseudoisothermal halo and the cuspy NFW halo
(Navarro, Frenk, \& White 1997).  It is straightforward 
to derive equivalent limits for other choices of halo models, but
most other models resemble one of these two in their essential features.

\subsubsection{Pseudoisothermal Halos}

Traditionally, galaxy rotation curves have been fit with pseudoisothermal 
halos.  These have constant density cores ($\rho = \rho_c$) out to a core
radius $R_c$, after which the profile rolls over to $\rho(r) \rightarrow r^{-2}$
in order to produce asymptotically flat rotation curves.  This functional form
is very effective at fitting rotation curves, though there is no guarantee
that this is what dark matter {\it should\/} do.

The dark matter distribution of the pseudoisothermal halo is
\begin{equation}
\rho_{\mathrm{ISO}}(r) = \frac{\rho_c}{1 + (r/R_c)^2}
\end{equation}
which gives rise to a rotation velocity
\begin{equation}
V_{\mathrm{ISO}}^2(r) = V_f^2 \left[ 1-\frac{R_c}{r} 
\tan^{-1}\left( \frac{r}{R_c} \right) \right].
\end{equation}
Only two of the parameters $(\rho_c, R_c, V_f)$ are independent, being
related by $V_f = R_c \sqrt{4 \pi G \rho_c}$.  
While it is natural to associate $V_f$ with
the observed $V_{flat}$, this need not be the case in detailed mass 
decompositions.  For nearly maximal disks the baryonic component is
often still important at the last measured point, leaving flexibility for 
$V_f \ne V_{flat}$.

The pseudoisothermal halo is a very flexible fitting function, and its
parameters are generally not well constrained by direct fits to rotation
curves.  The fit parameters are highly correlated,
particularly $R_c$ and \ML\ (Kent 1987).
As a result, one can trade one off against the other, leading to the long
standing disk-halo degeneracy.

The acceleration constraint helps break this degeneracy.  The centripetal
acceleration $a_{\mathrm{ISO}} = V_{\mathrm{ISO}}^2/r$ 
due to the halo (equation 13)
has a maximum at $r = 1.5 R_c$.  The constraint is then
\begin{equation}
a_{\mathrm{ISO}}^{\mathrm{max}} = 0.23 \frac{V_f^2}{R_c} <  \eta \amax.
\end{equation}

The acceleration constraint is fairly restrictive.  For $\amax = a_0$,
small core radii are disallowed for more rapidly rotating galaxies.  
Small core radii ($R_c < 1$ kpc) do not occur for typical
galaxy velocities ($V_f > 100$\kms) unless we invoke
implausibly small mass-to-light ratios.  Excessive halo velocities
($V_f > 250$\kms) are disallowed except for very large core radii
($R_c > 10$ kpc).  This is interesting as it can be quite difficult to
constrain $V_f$ in many cases where the halo contribution to the
velocity is still rising at the last measured point.

\subsubsection{NFW Halos} 

It is not obvious that one expects dark matter halos to be pseudoisothermal.
In numerical simulations of structure formation, cold dark matter (CDM)
is found to form halos with a rather different structure.  For example,
Navarro \etal\ (1997) find that their
simulated halos are reasonably well described by
\begin{equation}
\rho_{\mathrm{NFW}}(r) = \frac{\rho_i}{(r/R_s)(1+r/R_s)^2}.
\end{equation}
This gives rise to a circular velocity
\begin{equation}
V_{\mathrm{NFW}}^2(r) = V_{200}^2 \left[\frac{\ln(1+cu) - cu/(1+cu)}
{u[\ln(1+c)-c/(1+c)]}\right],
\end{equation}
where $c = R_{200}/R_s$ and $u=r/R_{200}$.
$R_{200}$ is the radius which encloses a density 200 times the critical density
of the universe (this is, crudely speaking, the virial radius), 
and $V_{200}$ is the circular velocity of the potential at $R_{200}$.

NFW halos do not have $\rho \propto r^{-2}$ except in transition
between two limits, $\rho \propto r^{-1}$ at small radii and 
$\rho \propto r^{-3}$ at large radii.  They do not predict flat rotation curves,
which must arise from a combination of disk plus halo.
The lack of a constant density core causes problems
in fitting the slowly rising rotation curves of dwarf and LSB galaxies
(Flores \& Primack 1994; Moore 1994; de Blok \etal\ 2001, 2003;
Swaters \etal\ 2003).   In HSB galaxies, some disk mass is required to
give flat rotation curves and to avoid excessively high concentrations,
but not too much in order to leave room for the central
cusp which requires substantial dark mass at small radii.

Maximum disks and NFW halos are mutually exclusive.
Even middle-weight disks can significantly modify the primordial halo
profile through adiabatic contraction (e.g., Sellwood 1999).  
We leave an investigation of this
point for future work, noting here only that the limits found here will
only become stronger after allowance for adiabatic contraction.

The maximum acceleration of an NFW halo occurs at $r = 0$.
Using equation~16 in the limit $u \rightarrow 0$, the acceleration limit is
\begin{equation}
a_{\mathrm{NFW}}^{\mathrm{max}} = \frac{1}{2} 
\left[\frac{V_{200}^2}{R_{200}}\right]
\frac{c^2}{\ln(1+c)-c/(1+c)} < \eta \amax.
\end{equation}
This can be simplified by noting that $V_{200} = R_{200} h$
(where $h = H_0/100 {\rm km}\,{\rm s}^{-1}\,{\rm Mpc}^{-1}$).
The limit obtained in this fashion is shown in Figure 8.
Quite a lot of parameter space is disallowed,
including nearly all of that which halos are predicted to occupy
by \LCDM\ (Navarro \etal\ 1997; see also McGaugh, Barker, \& de Blok 
2003) with the parameters fixed by WMAP (Bennett \etal\ 2003).

Also shown in Figure 8 are the limits placed by the data for low surface
brightness galaxies from de Blok \etal\ (2003) and Swaters \etal\ (2003).  
There is no requirement that all galaxies attain accelerations as high as
\amax, and most of these systems do not.  It is for this reason that the 
limits from some individual galaxies are more restrictive than the
limit imposed by equation~17, even when $\ML = 0$ is assumed.

The data require rather lower halo concentrations than are predicted by 
\LCDM.  This holds for {\it all\/} spirals, not just those of low surface brightness.
Indeed, NFW halos are more strongly excluded for the high mass halos
which we would presumably associate with $L^{\star}$ galaxies than for
lower mass dwarfs.
The only way to avoid this conclusion is to increase \amax\ by reducing
\ML.  The region of parameter space predicted by \LCDM\ is allowed for
$\amax > 10^4\kmskpc$.  This corresponds to $\ML <$ half popsynth
(see also Bell \etal\ 2003a).
Stellar mass-to-light ratios this small seem very unlikely considering the
large scatter in the mass discrepancy-acceleration relation and baryonic
Tully-Fisher relation that they cause (\S 3).  

\subsection{A Quantitative Relation Between Dark and Baryonic Mass}

The information contained in the data goes well beyond limits on the 
parameters of simple halo models.  The limits discussed in the previous 
section only make use of the scale \amax.  We can make greater use of
the correlation between the mass discrepancy ${\cal D}$ and
acceleration.  The functional form apparent in Fig.~5 provides a strong
constraint on the force produced by the dark matter at all radii.

The observed velocity is the sum, in quadrature, of the various components:
\begin{equation}
V^2(r) = \Vst^2(r) + \Vg^2(r) + \Vh^2(r).
\end{equation}
This sum of stars, gas, and dark matter
is one valid description of the the observed rotation curve.  
Another valid description is given by MOND fits:
\begin{equation}
V^2(r) = \mu^{-1}(x) \left[\ML \vst^2(r) + \Vg^2(r)\right].
\end{equation}
This is effectively just a scaling of the baryonic component: 
$V^2(r) = {\cal D}(x) \Vb^2(r)$.  That is, the same relation can be
derived in a purely Newtonian fashion from either Fig.~4 or 5 by fitting a 
function ${\cal D}(x)$ to the bottom panel
of Fig.~4 or ${\cal D}(y)$ to Fig.~5.  
This description is more compact, and depends only on the observed 
distribution of baryonic mass: the velocity due to the dark matter halo does 
not explicitly appear.
Rather than an arbitrary function of radius $\Vh(r)$, there is only a
single fit parameter (\ML) which must take a specific numerical value.

Equating these two expressions for $V^2(r)$
stipulates what is required of the dark matter halo:
\begin{equation}
\Vh^2(r) = \left[\mu^{-1}(x) -1\right] \left[\Vst^2(r) + \Vg^2(r)\right].
\end{equation}
This equation encapsulates the coupling between mass and light.
The observed distribution of baryons
is completely predictive of the distribution of dark matter.

There is one, and only one, degree of freedom which is not incorporated
in equation~20.  It assumes that the MOND mass to light ratios are
correct.  While this would certainly appear to be the case,
we can generalize equation~20 to account for the possibility that
the true mass-to-light ratio $\ML^{\mathrm{TRUE}}$
differs from the optimal mass-to-light ratio $\ML^{\mathrm{MOND}}$.
These two mass-to-light ratios are related by
\begin{equation}
\Q \equiv \frac{\ML^{\mathrm{TRUE}}}{\ML^{\mathrm{MOND}}}.
\end{equation}
We can therefore write, with complete generality,
\begin{equation}
\Vh^2(r) = \left[\Q^{-1} \mu^{-1}(x) -1\right] \Vst^2(r) 
+ \left[\mu^{-1}(x) - 1\right] \Vg^2(r).
\end{equation}
This equation makes use of the compact MOND description of rotation curves
to specify the distribution of dark matter.  It depends not at all on MOND
being correct as a theory, but only as a fitting function for rotation curves.

Equation~22 mathematically expresses the strong coupling long noted
between mass and light in disk galaxies.  It is valid for {\it any\/} choice of
mass-to-light ratio, as encapsulated by \Q, the only free parameter. 
As seen in \S 3, the most natural value of \Q\ is unity.  Large deviations from
unity will disagree with stellar population constraints, so $\Vh(r)$ is
well constrained.  

Some examples of dark matter rotation curves are shown in Fig.~9 
for $\Q = 1$.  The overall shape of $\Vh(r)$ is well specified once the
baryonic component has been subtracted.  
Modest deviations from circular motion are probably the cause of
some detailed features (e.g., the sharp discontinuity in NGC 4013
which no smooth model can hope to fit)
but some other ripples in the inferred dark matter distribution may be
real (e.g., the dip in NGC 1560 which is present in the baryonic
distribution and reproduced by the MOND fit:  Begeman \etal\ 1991).

The dark matter halos are a mixed bag.  Some are dominant at all radii,
especially in the dwarf galaxies, while others appear nearly hollow at
small radii and never completely dominate even at the last measured
point.  This occurs for some of the more massive, bulge-dominated
systems, reminiscent of the lack of dark matter in the centers
of elliptical galaxies (Romanowsky \etal\ 2003).  Overall, it is hard
to generalize.  Each halo has its own detailed distribution, as unique
as that of the baryons which reside within it.

While the overall shape of $\Vh(r)$ is well specified, it does not follow
that model halos fit to these data will be well constrained, even with
\Q\ held fixed.  Models like the
pseudoisothermal and NFW halos discussed above have degeneracies
between their parameters that can make rather different halos look very
similar over a finite range of radii.  Observations necessarily span a
finite range, and usually only a small fraction of the anticipated virial
radius of the halo.  Making and interpreting such
fits is left for future work, and should be approached with caution.  

Having a specific equation for $\Vh(r)$ is nevertheless of considerable
value.  Much of the apparent freedom in multi-component mass modeling 
results from consideration of
arbitrary values of \Q\ ($0 \le \Q \le \Q_{\mathrm{max}}$)
and the degeneracies inherent in halo models.
This is a problem with how we analyze the data, not with the data 
themselves.  Indeed, we may be doing ourselves a considerable disservice
by trying to force preconceived functional forms upon $\Vh(r)$, as
equation~22 contains more detailed information than can be expressed 
by simple halo models.  Moreover, we do expect
some modification of the initial dark matter halo in the process of
galaxy formation, so it is not obvious that fitting NFW halos to the
current dark mass distribution is even an appropriate procedure.
Nevertheless, it remains true (as seen in Fig.~8) that present-day
dark matter halos have considerably {\it less\/} mass at small radii 
than nominally expected for primordial NFW halos.

\subsection{A Constraint on Galaxy Formation Models}

The halo circular velocity curve $V_h(r)$ derived from the mass
discrepancy-acceleration relation provides a strong constraint on
theories of galaxy formation.  For a model to be an acceptable 
representation of a real galaxy, it must be able to successfully 
apply equation~22 with \Q\ as the only free parameter.  
That is, one needs to be able to take the detailed distribution 
of stars and gas in the model and find a plausible value of \Q\ 
which gives $V_h(r)$ consistent with that of the model halo.

To qualify for this test, a model must be sophisticated enough to 
give both a baryonic mass distribution and a rotation curve.
The baryonic mass distribution must be more complex than a
simple exponential disk; it should reflect the real variation in
bulge, stellar disk, and gas disk components.
While model dark matter halos have been investigated in great detail,
obtaining a robust prediction of the baryonic mass distribution has proven
rather more difficult.  This depends not only on the details of dissipational
collapse, but also on the star formation prescription necessary to distinguish
stellar and gaseous components.  Models which meet these requirements
are only beginning to appear (e.g., Adabi \etal\ 2003; 
Zavala \etal\ 2003; Robertson \etal\ 2004).

The model of Abadi \etal\ (2003) provides a good example of the significant
hurdles theory still faces.  These authors succeed in forming,
in a numerical simulation, something which photometrically resembles a 
real disk plus bulge system.  While appealing, this model fails the test
posed above.  As Abadi \etal\ (2003) show, the object's
rotation curve is far removed from its photometric equivalent, peaking
much too high at a very small radius.  To make a proper comparison with 
data, one needs a large ensemble of such models.  Consequently,
we are still far away from being able to claim that we have a satisfactory
explanation of disk galaxy rotation curves.

Indeed, the particular form of the coupling between baryonic and total 
mass is rather peculiar.  There is no obvious reason to expect such a
coupling, let alone one so strong that it can be attributed to a single
{\it effective\/} force law which appears to be universal in disks.
One can write the equations as above (see also Dunkel 2004), 
but this merely tells us {\it what\/} is required, not {\it why\/} it happens.  
By way of analogy, we might just as
well suppose that the solar system is really run by an inverse cube law.
It just looks like it operates under an {\it effective\/} inverse square law 
because there is dark matter arranged {\it just so}.  Such a situation in the
solar system would be considered an unacceptable fine-tuning problem.
The situation in galaxies is hardly better.

\section{Conclusions}

There is a strong relation between the distribution of
baryonic and total mass in disk galaxies.  The shape of the rotation curve
predicted by the observed distribution of baryons is homologous to the
observed rotation.  The mass discrepancy, defined as the ratio of
the gradients of the total to baryonic gravitational potential,
can be described by a simple function of centripetal acceleration:
\begin{displaymath}
\Phi'/\Phi_b' = {\cal D}(x),
\end{displaymath}
where $x = a/a_0$ and ${\cal D}(x)$ is the inverse of equation~6.
For $a > a_0 = 1.2 \times 10^{-10}\;{\rm m}\,{\rm s}^{-2}$, there is no
apparent need for dark matter.  For $a < a_0$, the amount of dark matter
increases systematically as acceleration declines.  

The empirical organization apparent in the data 
provides constraints on mass models
which have not been considered in traditional disk-halo modeling.  
Making full use of the information present in the data provides a method for 
estimating the stellar mass-to-light ratio of spiral disks.  These
mass-to-light ratios are optimal in that they
\begin{itemize}
\item minimize the scatter in the mass discrepancy-acceleration diagram
(Fig.~5),
\item minimize the scatter in the baryonic Tully-Fisher relation (Fig.~6), and
\item are consistent with the expectations of stellar population models (Fig.~7).
\end{itemize}
Indeed, the mass-to-light ratios are consistent not only with the mean value
anticipated for stellar populations, but they also reflect
the expected trends with color and the band-pass dependent scatter about 
the mean color-\ML\ relation.  It is hard to imagine that all this could be the
case unless these mass-to-light ratios are essentially correct.  This appears
to solve the long standing problem of the absolute stellar mass of galactic
disks.

Once the stellar mass of a disk is fixed, the distribution of dark matter
follows.  This can be expressed as a simple function of observable quantities:
\begin{displaymath}
\Vh^2(r) = \left[\Q^{-1} {\cal D}(x) -1\right] \Vst^2(r)
+ \left[{\cal D}(x) - 1\right] \Vg^2(r)
\end{displaymath}
(equation~22).  The dark matter distribution is completely specified
by the observed baryonic matter distribution.  This presents a serious
fine-tuning problem for any theory of galaxy formation.

The only freedom
available in the detailed distribution of dark matter inferred from observation
is encapsulated in a single parameter \Q.  This parameter allows for the
unlikely possibility that the actual mass-to-light ratios of stars differs
from the optimum value described here.  With this limited freedom, the relation
derived between baryonic and total mass specifies the dark
matter distribution with complete generality.  

Explaining how this strong coupling between mass and light arises is a 
major challenge for galaxy formation theory.
All we have done is restate, in a generalized Newtonian fashion, 
the well-established result that MOND fits rotation curves.
The question, of course, is what this means.
There are two independent issues here which are often mistakenly conflated.
One is the unconventional theory known as MOND. 
The second is the empirical regularity of the data. 
If the effective force law apparent in the data is not in fact MOND,
then we need to understand how it comes about in the context of dark matter. 

The possibilities are limited.  Either
\begin{itemize}
\item MOND is essentially correct, or
\item Dark matter results in MOND-like behavior in disk galaxies.
\end{itemize}
This is the same conclusion reached by Sanders \& Begeman (1994)
and McGaugh \& de Blok (1998b).  The improvement of the data since
that time only makes this result more clear.

If dark matter is correct, the tightness of the observed coupling between
baryons and dark matter implies a very strong regulatory mechanism. 
Within the context of dark matter, there seem to be two basic options.
Either
\begin{itemize}
\item the processes of galaxy formation lead to the observed coupling, or
\item there is a direct interaction between dark matter and baryons.
\end{itemize}  

In the first case, some combination of mundane astrophysical
effects (e.g., adiabatic contraction, mergers, feedback) are presumed 
to be the root cause of the coupling.  However,
no clear mechanism is known which has the required effect.
Indeed, it seems extremely unlikely that the
chaotic processes of galaxy formation can give such a highly
ordered result, much less the finely-tuned coupling between mass and light.

Alternatively, the strong coupling between dynamical and baryonic mass
might provide a hint about the nature of the dark matter itself.
Rather than interacting with baryons only through gravity, there may be
some direct interaction which results in the observed coupling.  
This amounts to a modification of the nature of the dark matter itself.
One obviously wishes to retain the successes of CDM on large scales,
but modify it to have the appropriate effect in individual galaxies.

Many modifications of dark matter have been discussed in recent years
(e.g., warm dark matter, self-interacting dark matter), motivated at
least in part by the difficulties posed by rotation curves.  Those proposed
modifications which ignore the baryons would not seem to help.  It is not 
adequate to change the dark matter properties in order to insert a soft core 
in dark matter halos; one must explain equation~22.

A mechanism which provides for the direct connection observed between
dark matter and baryons is unknown.  Ideas along this line
are only now being discussed (e.g., Piazza \& Marinoni 2003),
so it is too soon to judge whether they are viable.  Irrespective of which 
possibility might seem to hold the most promise, there is clearly some 
important physics at work which has yet to be understood.

\acknowledgements I would like to thank all the many people who have labored
so hard over the years to provide the wonderful data which have accumulated,
and the many others with whom I have discussed these issues.
Though there are too many to name here, I would in particular like to
acknowledge Kor Begeman, Albert Bosma, Adrick Broeils, Erwin de Blok,
Sandy Faber, Donald Lynden-Bell, Vera Rubin, Renzo Sancisi, Bob Sanders, 
Jerry Sellwood, Tjeerd van Albada, Thijs van der Hulst, and Marc Verheijen.
I also thank the referee and editor for their helpful comments.
The author acknowledges the support of the University of Maryland 
through a semester General Research Board award. 
The work of SSM is supported in part by NSF grant AST0206078 and
NASA ADP grant 02-0078-0050.


\begin{figure}  
\epsscale{1.0}
\plotone{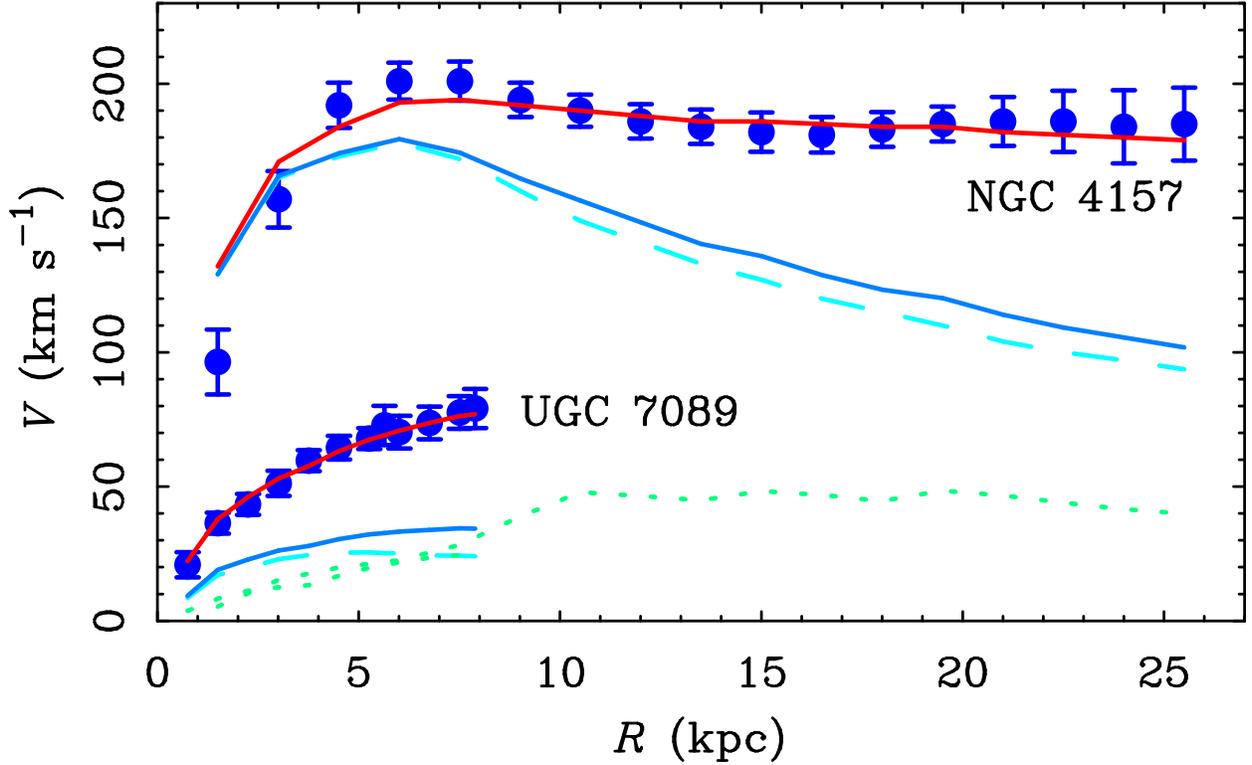}
\caption{Examples of the data for two galaxies from the study of
Sanders \& Verheijen (1998).  The points with error bars are the
measured rotation curves of NGC 4157 and UGC 7089.  The
dashed lines are the contribution to the rotation by the stars,
$\Vst(r)$, as determined from $K'$-band surface photometry.
The dotted lines are the contribution of the gas, $\Vg(r)$.  
Note that the gas distributions of these two very different galaxies
are rather similar for $r < 8$ kpc, so that the dotted lines nearly overlap.
The lower solid lines represent the total baryonic contribution,
$\Vb(r)$ (stars plus gas).  The upper solid lines are the MOND
fits.  These are, in effect, a version of $\Vb(r)$ scaled by the
MOND interpolation function $\mu(x)$.  Even if MOND is wrong
as a theory, this scaling provides an empirical relation between 
the distribution of observable baryons and that of the dark mass.
\label{Example}}
\end{figure}

\begin{figure}
\epsscale{1.0}
\plottwo{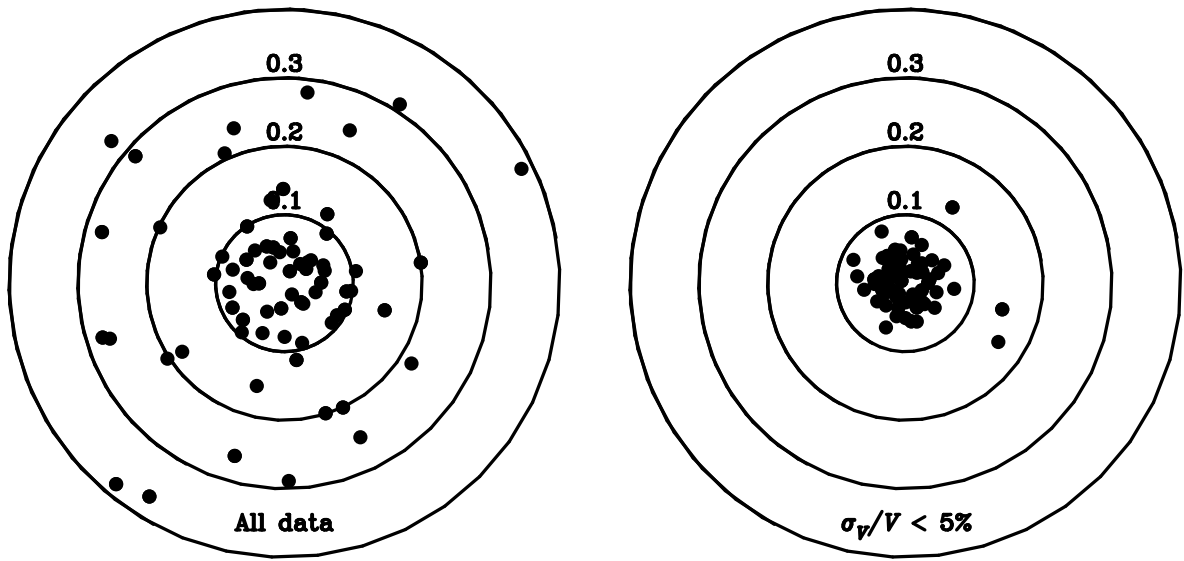}{f2b.eps}
\caption{The global rms residuals of MOND rotation curve fits.
The radial axis is the magnitude of the residual while the
azimuthal angle is assigned randomly to each object to distribute 
them about the plot.  Each point represents one galaxy.
In (a), all available data are used, a total of 74 galaxies.
In (b), the data are restricted to individual data points within
each galaxy with velocities measured to better than 5\% accuracy.
Fourteen galaxies have no data which satisfy this criterion,
leaving 60.  For these 60 galaxies, the one parameter MOND 
fits to the best data are very good. 
\label{rms}}
\end{figure}

\begin{figure}
\epsscale{1.0}
\plotone{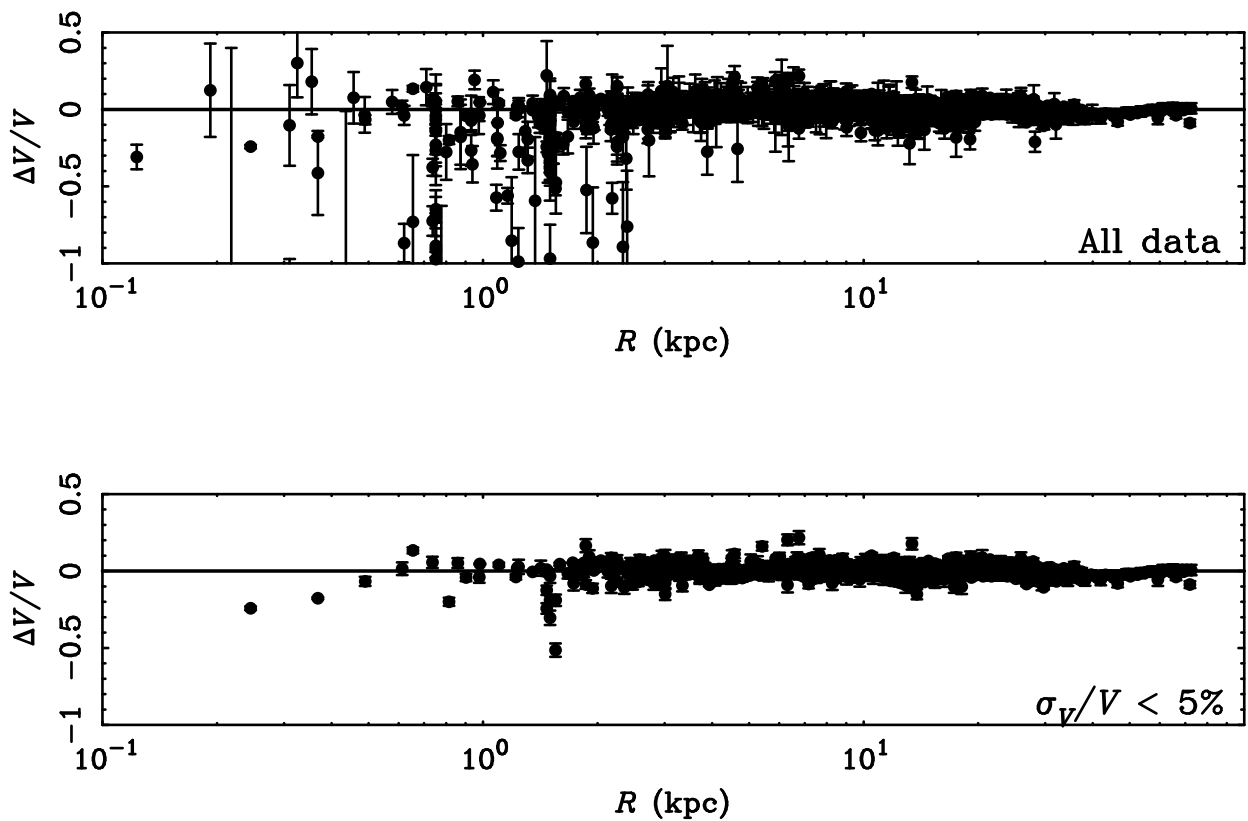}
\caption{The local residuals of MOND rotation curve fits.
Each point represents one resolved point within each galaxy; the data
for all galaxies are shown together.  The top panel shows the data for 
74 galaxies comprising a total of 1145 independent points.  
In the bottom panel, the data are restricted to individual data points within
each galaxy with velocities measured to better than 5\% accuracy, as in
Fig.~2.  A total of 736 independent points from 60 galaxies survive this cut. 
The places where the fits are imperfect tend to be where the measurements
are least accurate.  In general, the quality of the fits improves with the
quality of the data.  
\label{resid}}
\end{figure}

\begin{figure}
\epsscale{0.9}
\plotone{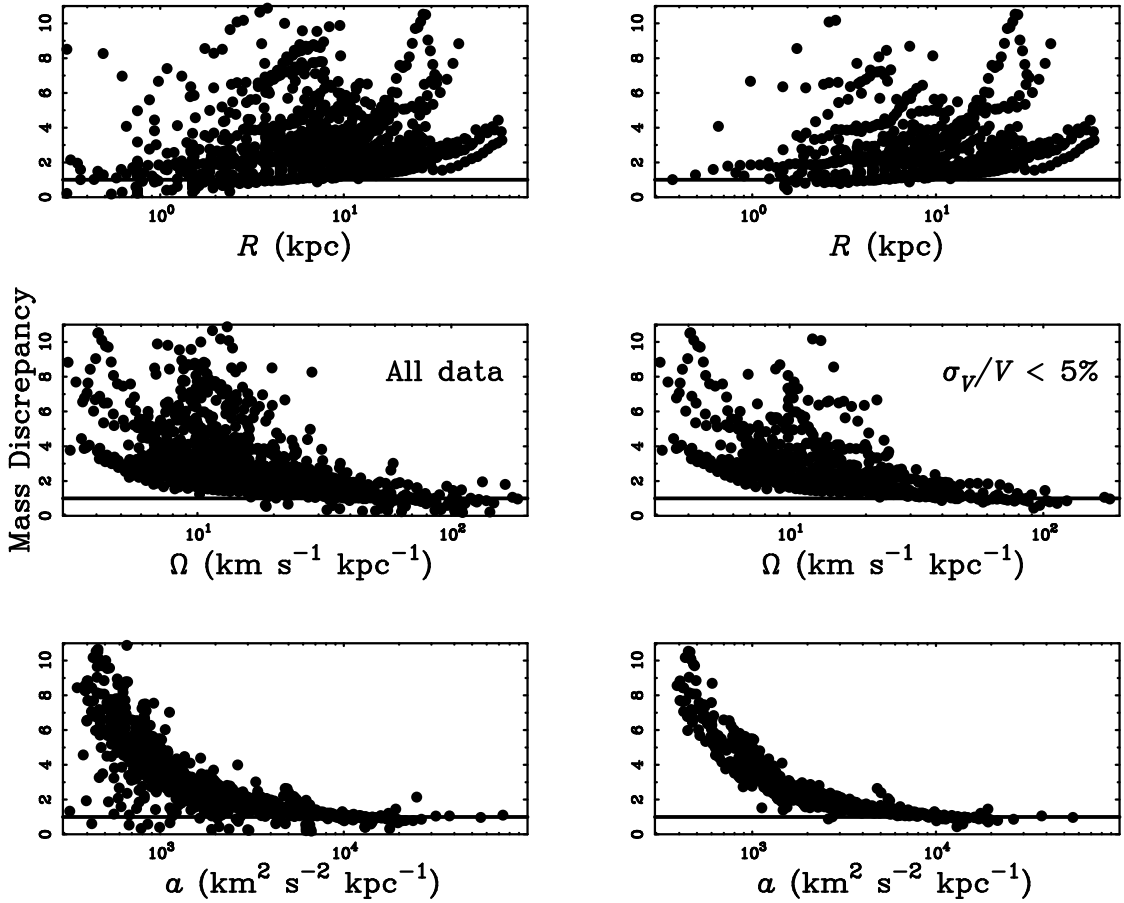}
\caption{The mass discrepancy ${\cal D} = V^2/V_b^2$ 
as a function of various observed scales:
radius (top panels); orbital frequency $\Omega = V/r$ (middle panels); and
centripetal acceleration $a = V^2/r$ (lower panels).  In the panels on the
left all data are shown; on the right are shown those data with high accuracy
($\sigma_V/V < 5\%$).  Each point represents one resolved point within each
galaxy, as in Fig.~3.  The horizontal line shows ${\cal D} = 1$ where there
is no mass discrepancy:  baryons suffice to explain the observed motions
here.  There is a clear organization of the data with
respect to acceleration.  The mass discrepancy only becomes apparent
(${\cal D} > 1$) below an acceleration scale of $a_0 \approx 
3700\;{\rm km}^2\,{\rm s}^{-2}\,{\rm kpc}^{-1}$
\label{MDRWA}}
\end{figure}

\begin{figure}
\epsscale{0.75}
\plotone{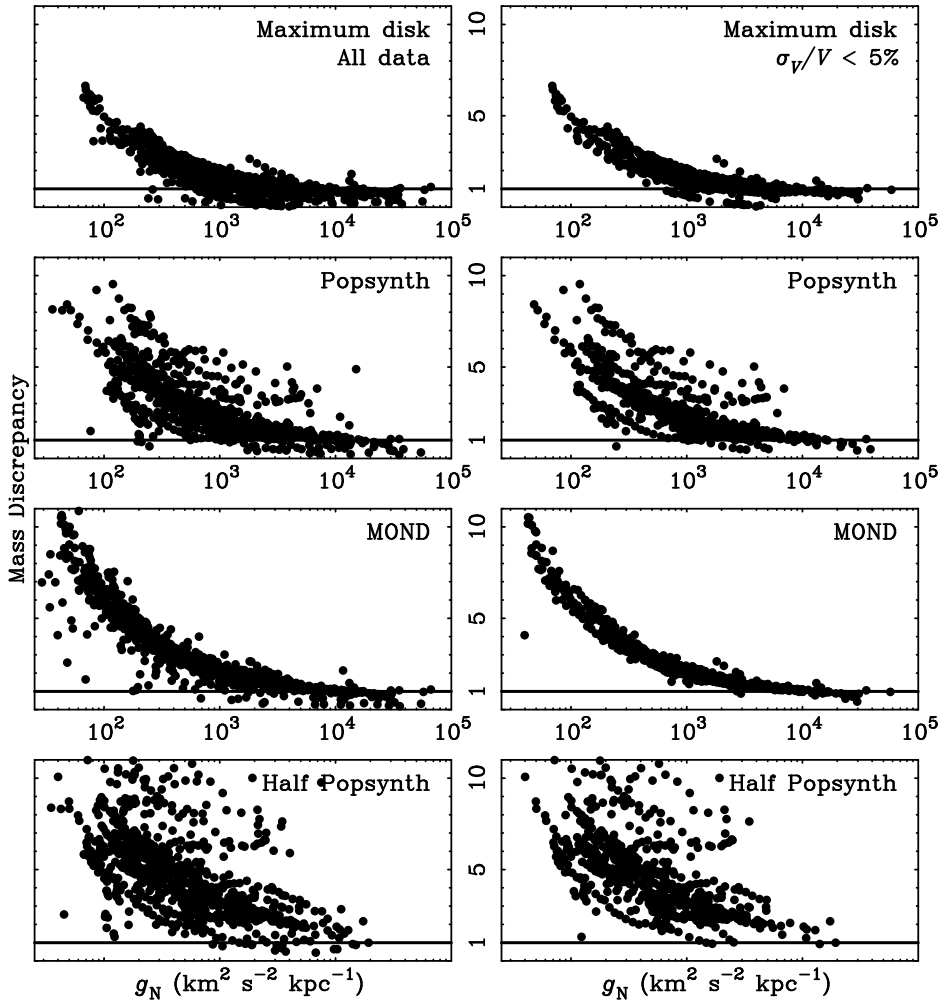}
\caption{The mass discrepancy as a function of the Newtonian 
acceleration $g_N = V_b^2/r$ predicted by the observed stars and gas
for various choices of mass-to-light ratio.  The topmost panels show the
result for maximum disk mass-to-light ratios.  The second row of panels
show the mass-to-light ratios indicated by the $B-V$ colors and the
\ML-color relation from the stellar population models
of Bell \etal\ (2003b).  The third row uses the MOND fit mass-to-light ratios
(Sanders \& McGaugh 2002).  The result of adopting a lightweight IMF is
illustrated in the bottom row by scaling the population synthesis 
mass-to-light ratio down by a factor of two.
The mass discrepancy is well correlated with the acceleration 
for any plausible (non-zero) choice of mass-to-light ratio.
The rotation curves of a few individual galaxies can be discerned in
the population synthesis rows, but the majority of the galaxies
are indistinguishable in all cases.
The scatter in this local relation is minimized for the MOND
mass-to-light ratios, making them optimal in this respect
even from a purely Newtonian perspective.
The intrinsic scatter of the relation appears to be very small.
\label{MDacc}}
\end{figure}

\begin{figure}
\epsscale{0.8}
\plotone{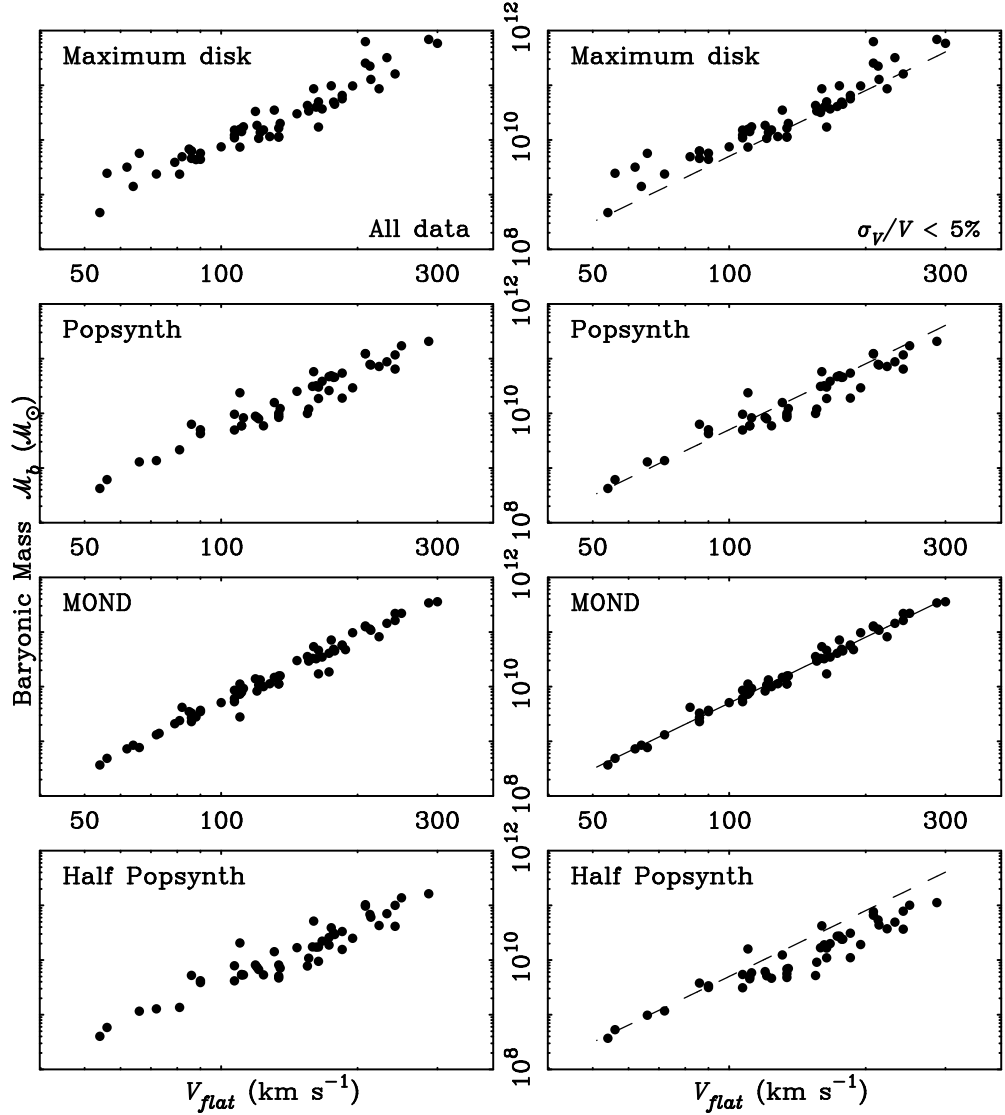}
\caption{The baryonic Tully-Fisher relation for various assumptions
about the stellar mass-to-light ratio, as in Fig.~5.  
The observed baryonic mass of each galaxy, stars plus gas, 
is plotted against the asymptotic flat rotation velocity.  
The scatter in this global relation is minimized for MOND mass-to-light
ratios.  The line fit to this case ($\mass_b = 50 V_{flat}^4$)
is shown in the right panels.  In the MOND panel it is shown as a solid 
line; in the other panels it is shown as a dashed line for reference.
Note that colors are only available for 50 of the 74 galaxies,
so not all cases appear in the popsynth panels.  
\label{BTF}}
\end{figure}

\begin{figure}
\epsscale{1.0}
\plotone{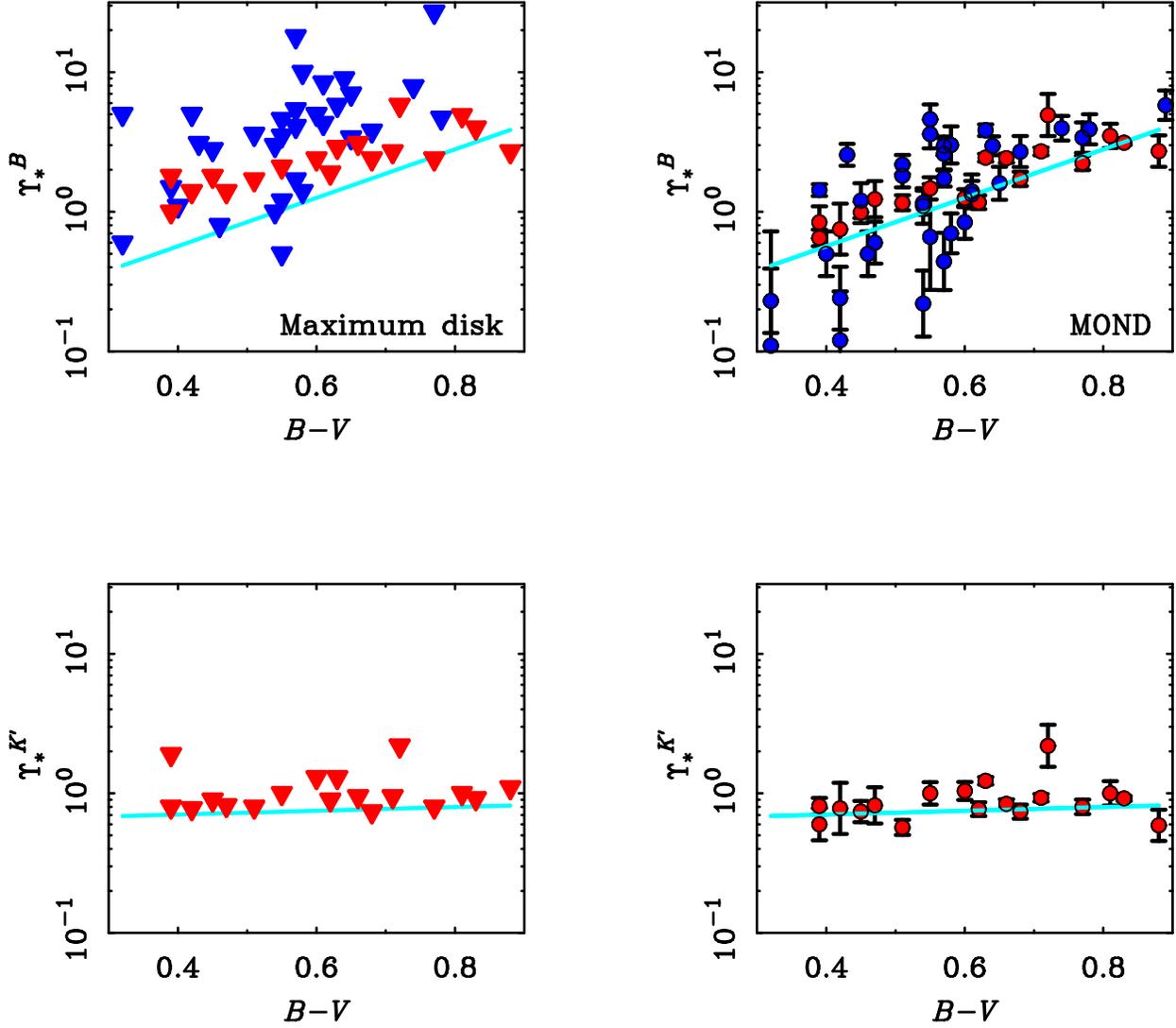}
\caption{The stellar mass-to-light ratios of disks in the $B$-band (top panels)
and $K'$-band (bottom panels) as a function of $B-V$ color for the cases of
maximum disk (left panels) and MOND (right panels).  Each point
represents one galaxy for which the requisite data exist.
The line represent the mean expectation of
stellar population synthesis models from Bell et al.\ (2003b).
These lines are completely independent of the data: neither the 
normalization nor the slope have been fit to the dynamical data.
\label{popsynth}}
\end{figure}

\begin{figure}
\epsscale{1.0}
\plotone{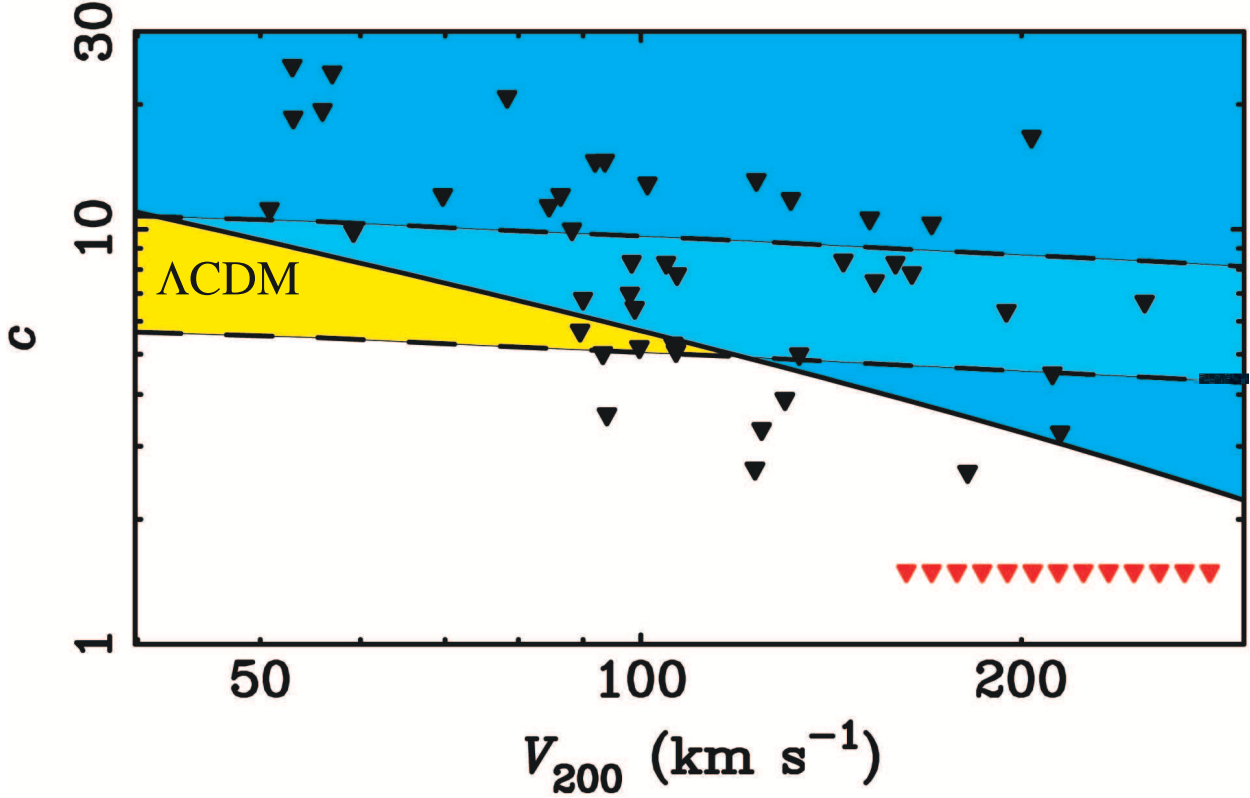}
\caption{Diagram showing the parameter space of NFW halos.  The dashed
lines enclose the region in which NFW halos are expected to reside for
the cosmological parameters derived from WMAP (Bennett \etal\ 2003).  
These lines are drawn for the $\pm 1 \sigma$ range of
scatter in concentration $\sigma_c = 0.14$ determined by
Weschler et al.\ (2002).  The blue region above the solid line is excluded 
by the maximum halo acceleration limit $\eta \amax$ (Brada \& Milgrom 1999).
This excludes most of the expected parameter space.  Also shown are the upper
limits on the concentrations of individual galaxies (points) measured in
the limit $\ML = 0$ by de Blok, McGaugh, \& Rubin (2001), de Blok \& Bosma
(2002), and Swaters et al.\ (2003).  Those objects which would fall outside
the boundaries of this plot because of exceedingly low concentrations
are plotted as the row of red points at lower right.  Individual galaxies need
not have halos which achieve the maximum acceleration limit, so such 
individual cases can be more restrictive (i.e., below the solid line).  
These are inevitably LSB galaxies with $a < a_0$ at all radii.
\label{NFWcV}}
\end{figure}

\begin{figure}
\epsscale{0.7}
\plotone{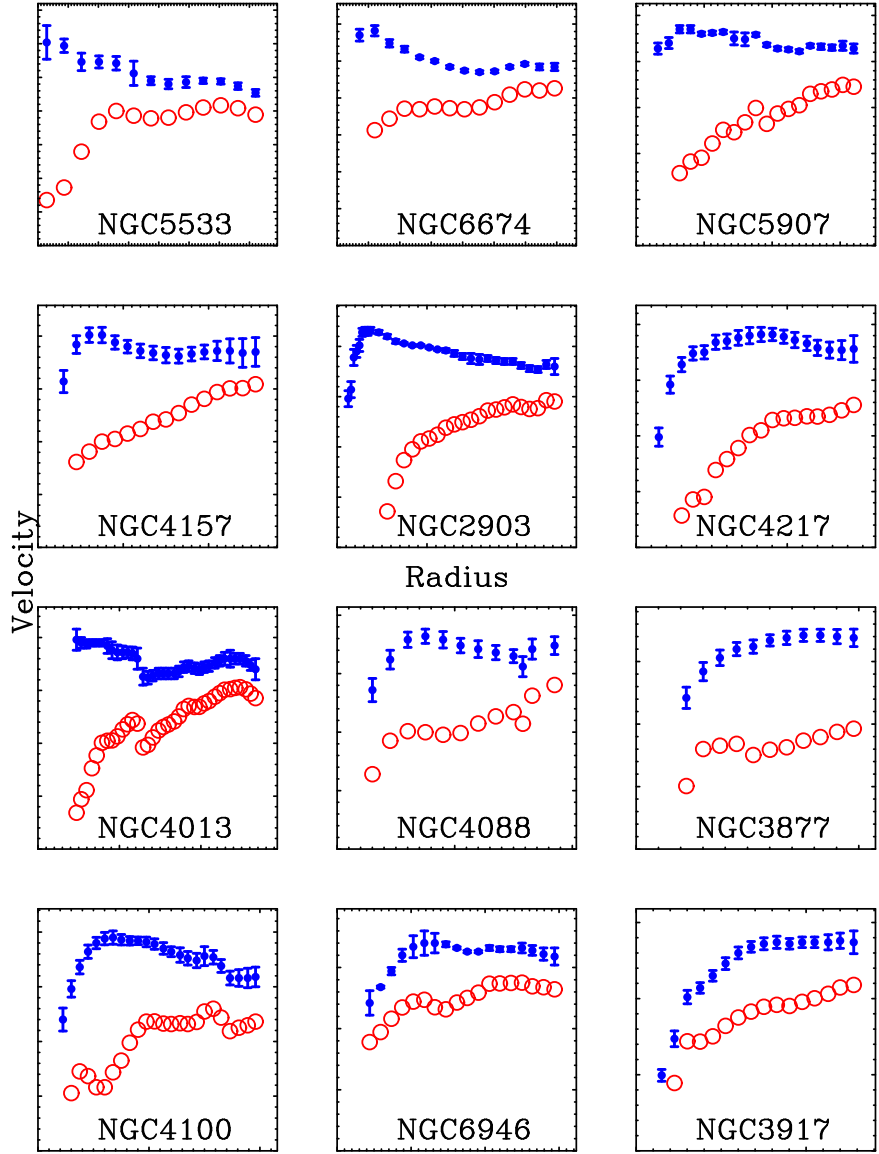}
\caption{The rotation curves $V(r)$ for individual galaxies with good data
(filled points) together with those inferred for the dark matter halo
$\Vh(r)$ after subtraction of the baryonic component (open points).  
The case of $\Q = 1$ is illustrated here.  All plots have (0,0) at the
bottom left corner.  Tick marks are 1 kpc along the abscissa and 
10\kms\ along the ordinate.  Galaxies are arranged in order of decreasing
velocity, from $V_{\mathrm{max}} \approx 300$\kms\ for NGC 5533 to 
$V_{\mathrm{max}} \approx 50$\kms\ for DDO 154.
There is a wide variety in $\Vh(r)$, ranging from nearly hollow
halos to those that dominate down to small radius.  
\label{mddmA}}
\end{figure}

\begin{figure}
\figurenum{9}
\epsscale{0.7}
\plotone{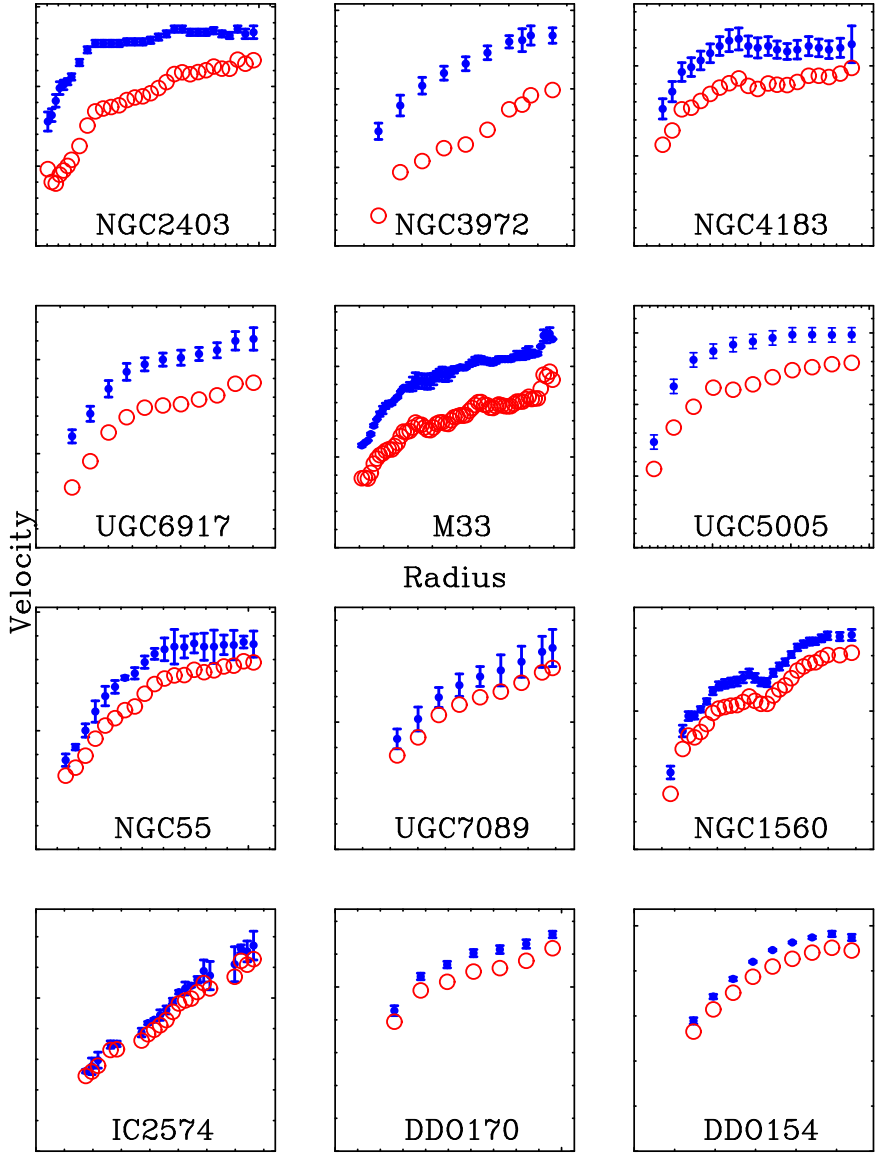}
\caption{continued.
\label{mddmB}}
\end{figure}

\end{document}